\documentclass{article}
\usepackage{graphicx} % Required for inserting images
\usepackage{amsmath}
\usepackage{bm}
\usepackage{tikz}
\usepackage[numbers]{natbib}
\usepackage{threeparttable}
\usepackage{amsfonts}
\usepackage{graphicx}
\usepackage{caption}
\usepackage{subcaption}
\usepackage{hyperref}
\usepackage{algorithm}
\usepackage{array}
\usepackage{booktabs}
\usepackage{pdflscape}
\usepackage{adjustbox}
\usepackage{subcaption} % Use the subcaption package for subfigures
\usepackage{svg}
\usepackage[margin=3.5cm]{geometry}
\tikzstyle{arrow1} = [black,thick,->,>=stealth]
\tikzstyle{arrow2} = [red, dashed,thick,->,>=stealth]
\usetikzlibrary{arrows, positioning, bayesnet}

\title{Compounding Effect of Harsh Climate and Societal Disruptions on Food Prices in Early Modern Europe}
\author{Emile Esmaili, Michael J. Puma, Francis Ludlow, Eva Jobbova}
\date{\today}

\begin{document}

\maketitle

\begin{abstract}

The complex interplay between famine, warfare, and climate constitutes a multifaceted and context-dependent relationship that has profoundly influenced human history, particularly in early modern Europe. This study advances the literature on climate-economy interactions by leveraging multi-scale statistical techniques to quantify the compounded effects of climate variability and socio-political factors on food prices, offering novel model-based insights into the historical dynamics of climate and economic systems.

Using Canonical Correlation Analysis (CCA), we investigate the influence of temperature fluctuations and drought severity on food prices across 14 European cities from 1565 to 1785. Our findings confirm a persistent negative relationship between temperature and food prices over the long term, while the relationship between drought severity and price dynamics appears positive yet inconsistent. Extending our analysis to higher-frequency patterns, we demonstrate that cold anomalies are strongly associated with food price that caused large-scale famines of the 1590s and 1690s. Likewise, we show that the severe and consecutive droughts of 1634–1636, coinciding with the Thirty Years' War, significantly amplified food price volatility, illustrating how climatic shocks can compound socio-economic and political crises. Furthermore, we identify years characterized by the simultaneous occurrence of extreme cold and drought as periods of heightened price instability, underscoring the compounded impact of concurrent climatic stressors on food prices during the early modern period.
\end{abstract}

\section{Introduction}

Many studies across disciplines have investigated the influence of climate change on different aspects of European history (see reviews by \cite{ljungqvistreview,degroot2018climate,haldon2018history,diaz2014some}). The German geographer Eduard Brückner (1862–1927) was a pioneer in studying multidecadal climate variability and its relationship to historical economic and societal conditions \cite{bruckner1895einfluss}. Since then, scholars have argued that climatic factors primarily affected grain price fluctuations on an interannual, but not longer term, basis (\cite{kelly2014a,kelly2013waning}). However, in addition to establishing that extreme droughts coincided with years of exceptionally high grain prices, \citet{esper2017environmental} identified a persistent and significant negative relationship between long-term, especially multidecadal, grain prices and temperature trends in Southern and Central Europe (i.e., colder climates corresponded to higher prices). Furthermore, \citet{esper2017environmental} emphasized that the relationship between climate and grain prices was not static but evolved over time, particularly during the Thirty Years War (1618–1648) when socio-political factors began to play a more prominent role. Similarly, \citet{camenisch2018weather} have suggested that in the late Middle Ages and early modern period, the link between climatic conditions and grain prices was stronger than that between weather/climate conditions and crop yields. The rationale provided is that grain prices reflect production across larger geographical areas and are thus less susceptible to local-scale growing conditions.

In terms of data focus, most studies have focused on grain prices due to their availability and direct correlation with harvest outcomes. However, our study diverges by focusing on broader food price indices derived from \cite{ALLEN2001411}, which, in addition to grains, include other commodities such as fish and meat—items comparatively overlooked in the literature. Regarding potential climate stressors, recent studies on the interaction between climate and society during the early modern period \cite{Ljungqvist2023,brazdil2023,Ljungqvistsignature} have concentrated primarily on temperature and hydroclimate variables (such as precipitation and drought indices), occasionally incorporating solar activity or soil moisture data. For hydroclimate, we use the Old World Drought Atlas \cite{owda}, which provides annual maps of tree-ring-reconstructed summer soil moisture throughout Europe and the Mediterranean during the Common Era. For temperature data, we use the ensemble mean of 2-meter air temperature from the Modern Era Reanalysis (ModE-RA) \cite{modeERA}, a global monthly paleo-reanalysis that covers the period from 1421 to 2008, blending information from transient atmospheric model simulations and observational data.

Recent scholarship by \citet{Ljungqvist2022significance, esper2017environmental,brazdil2023} has typically explored climate-price correlations on a pairwise basis to examine long-term trends without adjusting for confounding variables within the climate data (\citet{brazdil2024bohemia} employed Granger-causality but did not use multivariate regression models). These studies often employ spectral analysis or superimposed epoch analysis (SEA) to identify periodicities or patterns within univariate datasets. Our study aims to offer a more robust methodological framework for (1) establishing general causal relationships between climate and prices while accounting for spatial heterogeneity and (2) uncovering latent spatial and temporal patterns of maximum correlation between climate and prices through a multivariate approach. For (1), we utilize econometric frameworks widely used in economics to establish causal relationships in panel data (indexed by space and time), such as fixed-effects and mixed-effects models, which can capture both location and time heterogeneity. A noted limitation of these econometric models is their inherent monotonicity, which can not capture asymmetries (i.e., is drought worse than flood; cold worse than heat) or the potential time-varying nature of the relationships. While these econometric models help control for confounding factors and suggest potential causal relationships, it's important to note that they rely on strong assumptions about the nature of these relationships and the completeness of observed variables, which may not always hold in complex historical contexts.

For (2), we apply Canonical Correlation Analysis (CCA) \cite{hotellingrelations} and its sparse variant \cite{sparsemcca,parkhomenko2009sparse}, a multivariate statistical learning method designed to identify latent patterns of maximum correlation across multiple space-time matrices. CCA has found applications in fields such as neuroscience \cite{cca_neuro_review}, climate forecasting \cite{VANDENDOOL1994247,barnston92,wangagu2021}, natural language processing \cite{semanticsCCA}, computer vision \cite{nielsen2002multiset}, genetics \cite{witten2009penalized}, and acoustics \cite{wang2015unsupervised}. Notably, \citet{anderson2021food} have demonstrated the utility of this approach in a context similar to ours, using CCA to investigate the relationships among climate, food security, and conflict in a contemporary setting.

The resulting time-varying latent variables from the CCA will help to elucidate specific historical societal or climatic events that explain the patterns of maximum correlation between our variables. This study is the first to apply this multivariate statistical method to explore historical relationships between food prices and climate in the early modern period. Our work not only extends existing research, but also provides climate historians with a new set of tools for future investigations.

\section{Data and Methods}

\subsection{Data}

\subsubsection{Price Data}

We use yearly Consumer Price Indices (CPI) collected from \cite{ALLEN2001411} for 14 major European cities between 1565 and 1785. The cities included in the analysis are: Paris, London, Antwerp, Amsterdam, Gdansk, Lviv (Lwow), Warsaw, Vienna, Krakow, Madrid, Valencia, Augsburg, Leipzig, and Strasbourg. We winsorize the raw CPI values to 1 percent level (location-wise), detrend them, and standardize them. These preprocessing steps are essential for our analyses: winsorizing reduces the impact of extreme outliers that may distort the analysis, detrending removes long-term trends to focus on shorter-term fluctuations, and standardization ensures comparability across different locations by putting all series on a common scale.

\subsubsection{Climate data}

Our climate analysis focuses on two key variables: temperature and drought. For droughts, we use the Old World Drought Atlas\cite{owda}, a set of year-to-year maps of tree-ring-based summer soil moisture conditions (in the form of the self-calibrating Palmer Drought Severity Index) over Europe and the Mediterranean Basin during the Common Era. We populate a space-time matrix of $T$ years and $N$ locations, where each entry is the PDSI grid value at year $t$ around location $n$ within a 0.5 degree latitude and longitude box. 

For temperature we use the ensemble mean 2-meter air temperature from the Modern Era Reanalysis (ModE-RA) \cite{modeERA}. The Modern Era Reanalysis (ModE-RA) is a global monthly paleo-reanalysis covering the period between 1421 and 2008. To reconstruct past climate fields, an offline data assimilation approach is used, blending together information from an ensemble of transient atmospheric model simulations and observations. In the early period, ModE-RA utilizes natural proxies and documentary data, while from the 17th century onward instrumental measurements are also assimilated. We populate a second space-time matrix of shape $T \times N$ where every element is the closest grid point to every location's location. Amsterdam and Antwerp share the same fields as the grid is coarser for ModE-RA (1.8 degree $\times$ 1.8 degree). Note that, in order to weight the PDSI and temperature input fields according to their area on a sphere, each grid point is multiplied by $\sqrt{\cos{\phi}}$ where $\phi$ is the latitude. 

\subsection{Methods}

\subsubsection{General relationships between prices and climate}
To investigate the overall relationship between prices, temperature, and the PDSI, we first consider the following two-way fixed effects panel regression model:

\[
\text{CPI}_{it} = \beta_1 \text{temp}_{it} + \beta_2 \text{PDSI}_{it} + \gamma_i + \lambda_t + \varepsilon_{it}
\]

where: \(\text{CPI}_{it}\) is the dependent variable representing the Consumer Price Index for location \(i\) at time \(t\).\(\text{temp}_{it}\) and \(\text{PDSI}_{it}\) are independent variables (temperature and Palmer Drought Severity Index, respectively) for location \(i\) at time \(t\). \(\beta_1\) and \(\beta_2\) are the coefficients for the predictor variables \(\text{temp}_{it}\) and \(\text{PDSI}_{it}\), respectively. \(\gamma_i\) represents the location-specific fixed effects, capturing unobserved heterogeneity across units. \(\lambda_t\) represents the time-specific fixed effects, capturing unobserved heterogeneity across time periods. \(\varepsilon_{it}\) is the error term.
The heteroskedasticity robust standard errors are clustered at the location level following \citet{arellano87}. This approach accounts for potential correlation of errors within each location over time, ensuring more accurate inference in the presence of both heteroskedasticity and within-group error correlation. \\ 

We also specify a linear mixed-effects model to capture location-specific variation in the impact of climate variables on prices and to provide an additional check on the results from the fixed-effects model. The model is formulated as follows:

\[
\text{CPI}_{it} = \beta_0 + \beta_1 \text{temp}_{it} + \beta_2 \text{PDSI}_{it} + (u_i + u_{i1} \text{temp}_{it} + u_{i2} \text{PDSI}_{it}) + \epsilon_{it}
\]

where:
 \(\text{CPI}_{it}\) is the Consumer Price Index for location \(i\) at time \(t\). \(\beta_0\) is the fixed intercept. \(\beta_1\) and \(\beta_2\) are the fixed effects coefficients for temperature (\(\text{temp}_{it}\)) and PDSI (\(\text{PDSI}_{it}\)), respectively. \(u_i\) is the random intercept for location \(i\). 
 \(u_{i1}\) and \(u_{i2}\) are the random slopes for temperature (\(\text{temp}_{it}\)) and PDSI (\(\text{PDSI}_{it}\)) for location \(i\).
 \(\epsilon_{it}\) is the error term.
The model is estimated using Restricted Maximum Likelihood\cite{reml}, and cluster-robust (CR2\cite{cr2}) standard errors are reported. These methods are chosen to handle the hierarchical data structure and account for within-location error correlation, ensuring robust inference 

\subsubsection{Canonical Correlation Analysis}

Canonical Correlation Analysis (CCA)\cite{hotellingrelations} is a statistical method for identifying linear combinations of two sets of variables that have maximum correlation with each other. We apply CCA to uncover patterns of maximum correlation in our space-time datasets of prices and climate variables.

Given two sets of variables:
\[
\mathbf{X} = (X_1, X_2, \ldots, X_p)^\top \quad \text{and} \quad \mathbf{Y} = (Y_1, Y_2, \ldots, Y_q)^\top,
\]
where \(\mathbf{X}\) is a \(p\)-dimensional random vector and \(\mathbf{Y}\) is a \(q\)-dimensional random vector, CCA finds linear combinations:

\[
U = \mathbf{a}^\top \mathbf{X} = a_1 X_1 + a_2 X_2 + \cdots + a_p X_p,
\]
\[
V = \mathbf{b}^\top \mathbf{Y} = b_1 Y_1 + b_2 Y_2 + \cdots + b_q Y_q,
\]
such that the correlation between \(U\) and \(V\) is maximized.

The canonical correlations \(\rho_k\) are the square roots of the eigenvalues \(\lambda_k\) of the matrix:

\[
\Sigma_{XX}^{-1/2} \Sigma_{XY} \Sigma_{YY}^{-1} \Sigma_{YX} \Sigma_{XX}^{-1/2}.
\]

The canonical vectors are obtained by solving:

\[
\mathbf{a}_k = \Sigma_{XX}^{-1} \Sigma_{XY} \mathbf{b}_k,
\]
\[
\mathbf{b}_k = \Sigma_{YY}^{-1} \Sigma_{YX} \mathbf{a}_k,
\]
where \(\mathbf{a}_k\) and \(\mathbf{b}_k\) are the canonical vectors for the \(k\)-th pair of canonical variables.

The canonical variables \(U_k\) and \(V_k\) are then given by:

\[
U_k = \mathbf{a}_k^\top \mathbf{X}, \quad V_k = \mathbf{b}_k^\top \mathbf{Y}.
\]

These canonical variables represent the linear combinations of our original variables that exhibit maximum correlation. By analyzing their behavior over time, we can infer patterns of maximum correlation between prices and each climate variable. 

\subsubsection{Sparse Multiple Canonical Correlation Analysis}
In order to analyse CPI, PDSI, and temperature matrices all at once, we use Sparse Multiple CCA (SMCCA)\cite{sparsemcca} to account for sparsity and multiple datasets. We will use Sparse MCCA on all three data matrices (CPI, temperature, and PDSI), and the entire grids instead of location-specific fields, making temperature and PDSI matrices very high-dimensional, and therefore requiring a sparse method for high-dimensional data.

Sparse Canonical Correlation Analysis (SCCA)\cite{parkhomenko2009sparse,witten2009penalized} can be extended to perform integrative analysis on more than two datasets with features on a single set of samples. When dealing with $K$ datasets, denoted as $X_1, X_2, \ldots, X_K$, where each dataset $X_k$ contains $p_k$ variables (features), and assuming that each variable has a mean of zero and a standard deviation of one, we can use a generalized CCA approach for multiple datasets.

The single-factor multiple-set CCA criterion involves finding vectors $w_1, w_2, \ldots, w_K$ that maximize the sum of pairwise covariances between datasets:

\[
\text{maximize} \quad \sum_{i < j} w_i^\top X_i^\top X_j w_j
\]

subject to the normalization constraints:

\[
w_k^\top X_k^\top X_k w_k = 1, \quad \forall k.
\]

Here, $w_k \in \mathbb{R}^{p_k}$ is the weight vector for dataset $X_k$.

To introduce sparsity, we can modify the multiple-set CCA criterion by imposing sparsity constraints. In the sparse multiple-set CCA (sparse mCCA), we assume that $X_k^\top X_k = I$ for each $k$. The criterion for sparse mCCA becomes:

\[
\text{maximize} \quad \sum_{i < j} w_i^\top X_i^\top X_j w_j
\]

subject to:

\[
\|w_i\|_2 \leq 1, \quad P_i(w_i) \leq c_i, \quad \forall i,
\]

where $\|w_i\|_2$ represents the Euclidean norm (ensuring normalization) and $P_i(w_i)$ represents a sparsity constraint with a threshold $c_i$ for each weight vector $w_i$.

\section{Results}

\subsection{Long-term general relationships between climate and food prices}

\begin{figure}[htbp]
    \centering
    \includegraphics[width=0.5\linewidth]{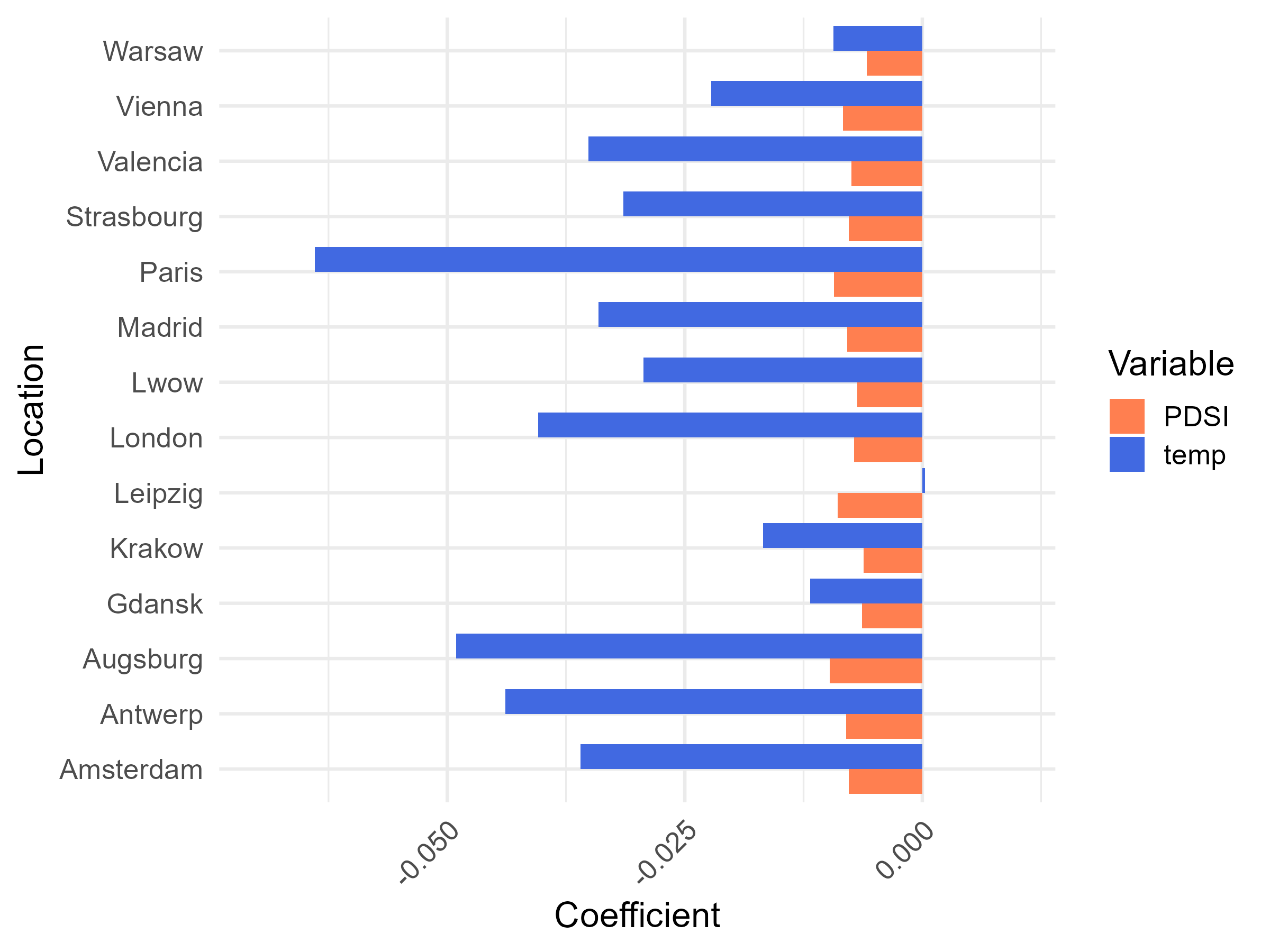}
    \caption{Coefficients from the linear mixed-effects model with random slopes on PDSI and temperature}
    \label{fig:lmercoefs}
\end{figure}

We first investigate long-run statistical relationship in the data between prices, temperature, and drought.  The analysis consists of panel and mixed-effects regressions, along with the standard Ordinary Least Squares (OLS) specification. The results in table \ref{tab:regs} indicate a statistically significant (p$<0.05$) negative relationship between CPI and temperature, and very little to no relationship between PDSI and CPI (the models disagree on the significance but the coefficient is very small and negative in all cases). Our results agree with a study by \citet{Ljungqvist2022significance} claiming that high grain prices were mainly associated with low temperatures, and inconsistently with droughts. We also report the location-specific coefficients from the mixed effects model in figure \ref{fig:lmercoefs}, which suggests some variability on the response from temperature but none from the PDSI.

\begin{table}
\caption{Dep. Var. CPI}
\begin{center}
\begin{threeparttable}
\begin{tabular}{l c c c}
\hline
 & OLS & Fixed-Effects & Mixed-Effects \\
\hline
Temperature & $-0.0268 \; (0.0086)^{**}$ & $-0.0275 \; (0.0114)^{*}$ & $-0.0302 \; (0.0104)^{**}$  \\
PDSI        & $-0.0073 \; (0.0037)^{*}$  & $-0.0018 \; (0.0031)$     & $-0.0077 \; (0.0020)^{***}$ \\
\hline
N           & $3094$                     & $3094$                    & $3094$                      \\
\hline
\end{tabular}
\begin{tablenotes}[flushleft]
\scriptsize{\item $^{***}p<0.001$; $^{**}p<0.01$; $^{*}p<0.05$. \ For the OLS: Newey-West santard errors are reported. For the Panel: Newey West standard errors are clustered at the location level. For the Mixed-Effects: CR2 robust standard errors are reported}
\end{tablenotes}
\end{threeparttable}
\label{tab:regs}
\end{center}
\end{table}

\subsection{Maximum Correlation patterns}

We now turn to the Canonical Correlation Analysis (CCA). We run two CCAs. One for the CPI matrix and the PDSI matrix, and one for the CPI matrix and the Temperature matrix. We then run a Sparse Multiple CCA on all three data matrices.

\subsubsection{Prices and Droughts}

\begin{figure}[htbp]
    \centering
    \includegraphics[width=0.6\linewidth]{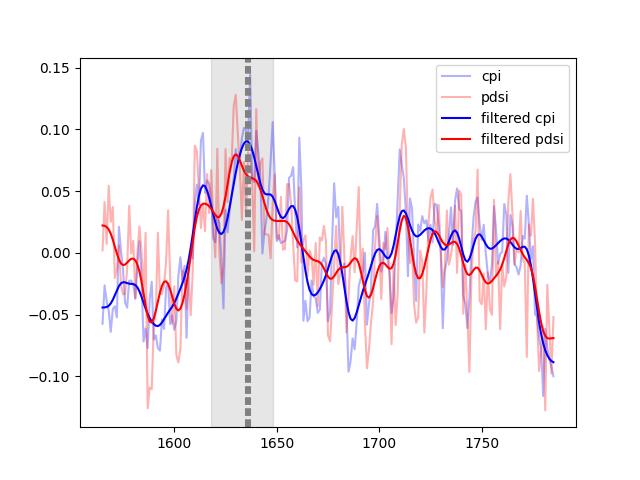}
    \caption{Canonical variables for the first component in raw and gaussian-filtered ($\sigma = 3$) form. The gray area is the Thirty-Years War (1618-1648) and the dashed lines are years 1634,1635,1636.}
    \label{fig:cca_variables_pdsi}
\end{figure}

\begin{figure}[htbp]
    \centering
    \includegraphics[width=\linewidth]{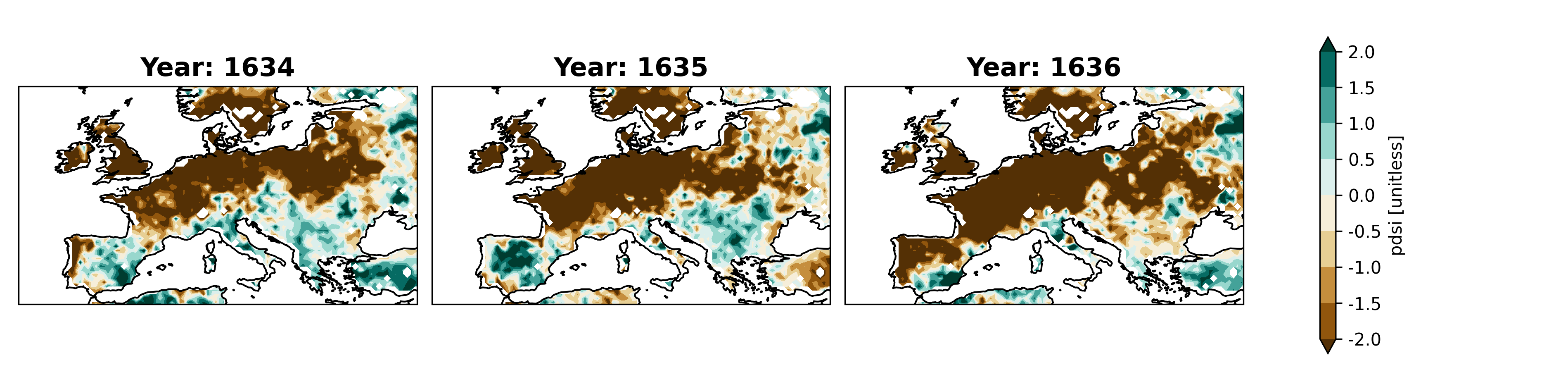}
    \caption{Main years of drought detected by the CCA.}
    \label{fig:key_droughts}
\end{figure}

The correlation between the first canonical variables is 0.6 (p$<$0.01 according to Wilk's lambda test). The canonical variables in figure \ref{fig:cca_variables_pdsi} suggest the pattern of maximum correlation between Prices and PDSI are explained during The Thirty Years' War (1618-1648), with noticeable spikes around 1635-37. Those years are among some of the most severe droughts documented in the 17th century \cite{owda} and as seen in the PDSI raw data in figure \ref{fig:key_droughts}. Our analysis statistically characterizes that period with a shock across all locations between prices and PDSI. In other words, severe droughts amplified the toll of the Thirty Years' War on food prices.
% The raw data in figure \ref{fig:cpi_key_dates_pdsi} confirms an increase in food prices in most locations except in Paris and Amsterdam.

Our study brings quantitative confirmation to scholarly work linking climate and food security: poor harvests often resulted from drought\cite{ljungqvist2022climatic}, and the outcome of grain harvests had a direct impact on grain prices and subsequently led to food shortages and high prices \cite{brazdil2023,brazdil2013droughts}, and potentially famines \cite{esper2017environmental}. 

% \begin{figure}
%     \centering
%     \includesvg[width=0.99\linewidth]{weights_map_pdsi.svg}
%     \caption{Weights of the first canonical variables for both PDSI and CPI matrices}
%     \label{fig:weights_map_pdsi}
% \end{figure}

% \begin{figure}[htbp]
%     \centering
%     \includesvg[width=0.99\linewidth]{cpi_key_dates_pdsi.svg}
%     \caption{CPI time-series during the Thirty-Years War. 1634-1638}
%     \label{fig:cpi_key_dates_pdsi}
% \end{figure}

\paragraph{Inferring directions of compounding effects}

\begin{figure}[htbp]
    \centering
    \begin{subfigure}[b]{0.45\textwidth}
        \centering
        \begin{tikzpicture}
            \node[latent, minimum width=1.2cm, minimum height=1cm] (War) {30YWar};
            \node[latent, right=of War, minimum width=1.2cm, minimum height=1cm] (CPI) {CPI};
            \edge {War} {CPI};
            \plate{plate1}{(War)(CPI)}{};
            \node[latent, above=0.7cm of plate1] (PDSI) {Drought};
            \edge {PDSI} {plate1};

        \end{tikzpicture}
        \caption{PDSI influencing War $\rightarrow$ CPI}
        \label{fig:dag1}
    \end{subfigure}
    \begin{subfigure}[b]{0.45\textwidth}
        \centering
        \begin{tikzpicture}
            \node[latent, minimum width=1.2cm, minimum height=1cm] (PDSI) {Drought};
            \node[latent, right=of PDSI, minimum width=1.2cm, minimum height=1cm] (CPI) {CPI};
            \edge {PDSI} {CPI};
            \plate{plate1}{(PDSI)(CPI)}{};
            \node[latent, above=0.7cm of plate1] (War) {30Y War};
            \edge {War} {plate1};
        \end{tikzpicture}
        \caption{War influencing PDSI $\rightarrow$ CPI}
        \label{fig:dag2}
    \end{subfigure}
    % \begin{subfigure}[b]{0.45\textwidth}
    %     \centering
    %     \begin{tikzpicture}
    %         \node[latent, minimum width=1.2cm, minimum height=1cm] (Temp) {Temp};
    %         \node[latent, right=of Temp, minimum width=1.2cm, minimum height=1cm] (Famine) {Famine};
    %         \node[latent, right=of Famine, minimum width=1.2cm, minimum height=1cm] (CPI) {CPI};
    %         \edge {Temp} {Famine};
    %         \edge {Famine} {CPI};
            
    %     \end{tikzpicture}
    %     \caption{Trivial: Temperature causing famines leading to price movements}
    %     \label{fig:dag3}
    % \end{subfigure}
    % \begin{subfigure}[b]{0.45\textwidth}
    %     \centering
    %     \begin{tikzpicture}
    %         \node[latent, minimum width=1.2cm, minimum height=1cm] (Temp) {Temp};
    %         \node[latent, right=of Temp, minimum width=1.2cm, minimum height=1cm] (CPI) {CPI};
    %         \edge {Temp} {CPI};
    %         \plate{plate1}{(Temp)(CPI)}{};
    %         \node[latent, above=0.7cm of plate1] (Famine) {Famine};
    %         \edge {Famine} {plate1};
    %     \end{tikzpicture}
    %     \caption{\textbf{Non-trivial:} Famine aggravating the impact of weather on prices}
    %     \label{fig:dag4}
    % \end{subfigure}

    \caption{DAGs showing different relationships among War, drought (PDSI) and prices (CPI).}
    \label{fig:DAGs}
\end{figure}
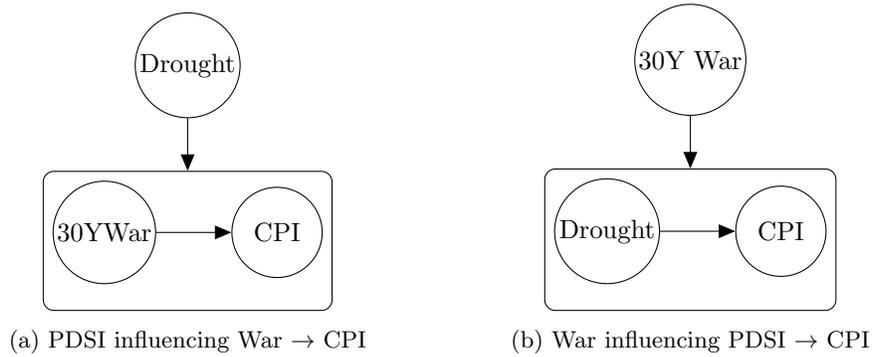

The previous analyses highlight a strong compounding effect of climate conditions and societal events (war and droughts) that reinforce price movements. We wish to understand how climate and socioeconomic factors compounded each other to influence price increases. We summarize these different relationships in figure \ref{fig:DAGs}.
% For (2), the more intuitive causal direction is that cold weather played a significant role in driving up prices and contributing to the emergence of the pan-European famines. We want to test the 'non-trivial' direction: the aggravation of the effects of cold weather on prices by these famines.\\

% We are interested in the relationships described by \ref{fig:dag2} and \ref{fig:dag4} 

While it is more reasonable to assume that the droughts reinforced the adverse impact of the War on prices, we also want to test whether the War had an impact on the drought/price relationship. The empirical strategy is carried out using the following specification:
\[
\rho(\text{CPI},climate)_{it} = \alpha_i + \beta_1 \text{TY\_war}_{t} + \beta_2 \text{famines}_{t} + \beta_3 CPI_{it} + \beta_4 climate_{it} + \varepsilon_{it}
\]

\noindent where: \(\rho(\text{CPI},climate)_{it}\) is the dependent variable representing the correlation between CPI and the climate variable, either temperature or PDSI, for location \(i\) at time \(t\), computed over different rolling windows (10Y, 15Y, 20Y). \(\alpha_i\) represents the individual-specific fixed effects (capturing unobserved heterogeneity across locations). \(\text{TY\_war}_{it}\) is an indicator variable that equals 1 for all years \(t\) during the Thirty-Years' War, and 0 otherwise.
 \(\text{famines}_{it}\) is an indicator variable that equals 1 if famine events occurred at time \(t\), and 0 otherwise. \(\varepsilon_{it}\) is the idiosyncratic error term.
 
The different regression analyses summarized in table \ref{tab:regs_cor_PDSI} suggest a consistent effect of the War on the accentuation of the drought/higher prices relationship (i.e. negative CPI/PDSI correlation). 
% Similarly, it is hypothesized that cold weather was responsible for the pan-European Famines of the 1590s and 1690s. Yet in table \ref{tab:regs_cor_temp} we estimate a robust and consistently significant effect of famines on the how cold weather led to higher prices (i.e. negative CPI/Temperature correlation). 
% In both cases we deduce that these compounding effects were two-way effects that fed on each other.

\begin{table}
\caption{Dep. Var. : PDSI/CPI Correlation}
\begin{center}
\begin{threeparttable}
\begin{tabular}{l c c c}
\hline
 & 10Y corr & 15Y corr & 20Y corr \\
\hline
TY\_war & $-0.079 \; (0.012)^{***}$ & $-0.076 \; (0.011)^{***}$ & $-0.063 \; (0.010)^{***}$ \\
famines & $0.006 \; (0.013)$        & $-0.015 \; (0.010)$       & $-0.023 \; (0.009)^{**}$  \\
CPI     & $0.016 \; (0.022)$        & $0.063 \; (0.019)^{***}$  & $0.043 \; (0.017)^{*}$    \\
PDSI    & $-0.001 \; (0.003)$       & $-0.003 \; (0.002)$       & $-0.004 \; (0.002)$       \\
\hline
N       & $2828$                    & $2828$                    & $2828$                    \\
\hline
\end{tabular}
\begin{tablenotes}[flushleft]
\scriptsize{\item $^{***}p<0.001$; $^{**}p<0.01$; $^{*}p<0.05$. \  Newey-West standard errors are clustered at the location level.}
\end{tablenotes}
\end{threeparttable}
\label{tab:regs_cor_PDSI}
\end{center}
\end{table}

\subsubsection{Prices and Cold Weather}

\begin{figure}
    \centering
    \includegraphics[width=0.6\linewidth]{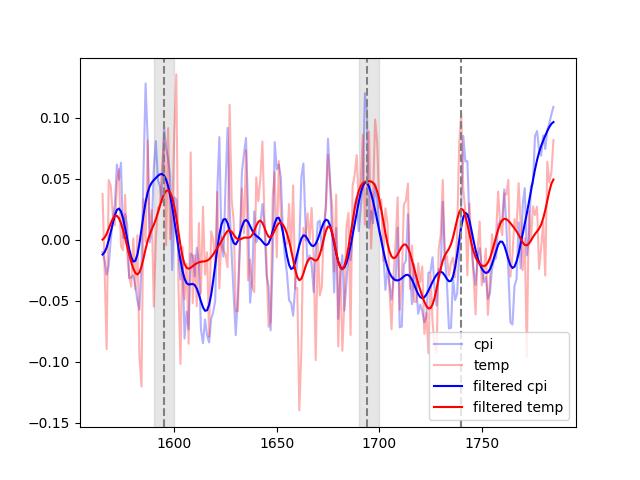}
    \caption{Canonical variables for the first component, raw and gaussian-filtered ($\sigma = 3$)}
    \label{fig:cca_variables_temp}
\end{figure}

\begin{figure}
    \centering

    % Subfigure 1
    \begin{subfigure}[b]{0.99\linewidth}
       \caption{1590s}
        \includegraphics[width=\linewidth]{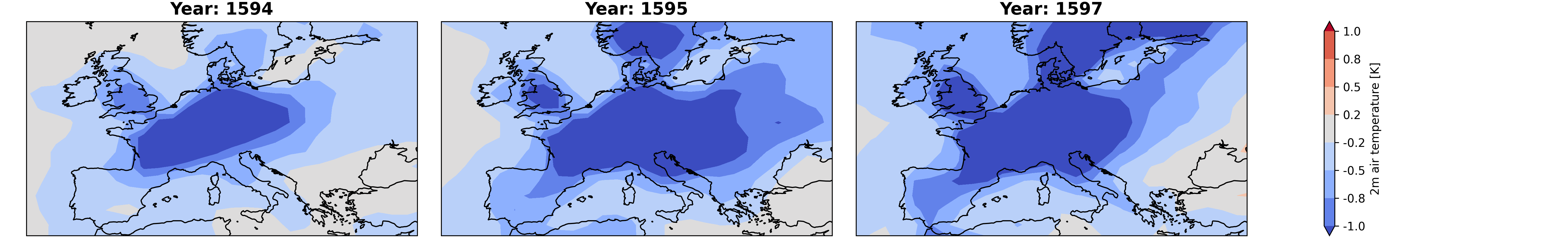}
     
        \label{fig:sub1}
    \end{subfigure}
    \hfill % Adds horizontal space between subfigures

    % Subfigure 2
    \begin{subfigure}[b]{0.99\linewidth}
            \caption{1690s}
        \includegraphics[width=\linewidth]{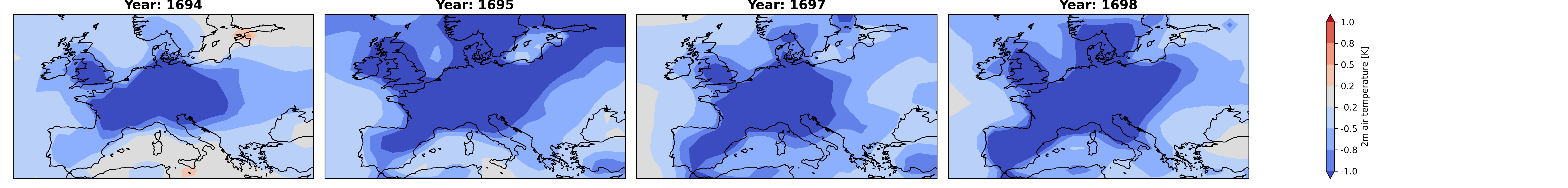}

        \label{fig:sub2}
    \end{subfigure}
    \hfill % Adds horizontal space between subfigures

    % Subfigure 3
    \begin{subfigure}[b]{0.99\linewidth}
            \caption{1700s}
        \includegraphics[width=\linewidth]{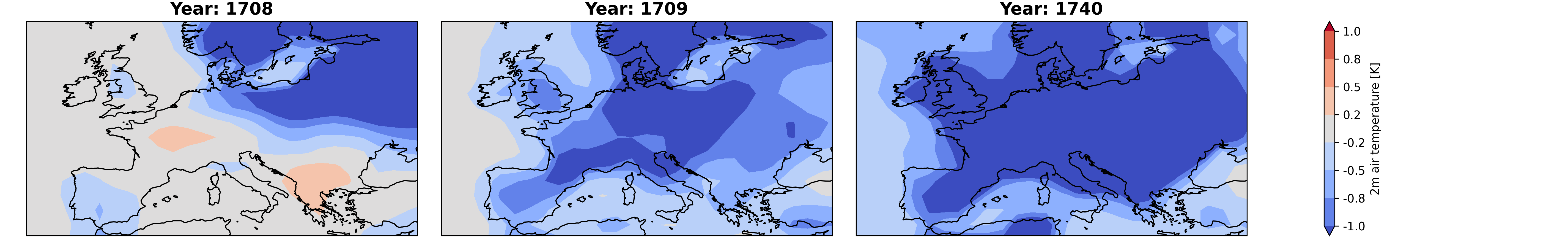}
        \label{fig:sub3}
    \end{subfigure}

    \caption{Years of cold temperature from ModE-RA (yearly average) in the 1590s, 1690s, 1700s}
    \label{fig:key_temperatures_decades}
\end{figure}

The first canonical correlation is 0.49 (p$<$0.01 according to Wilk's lambda test). The variables' time series in figure \ref{fig:cca_variables_temp} suggest the latent patterns of maximum correlation between prices and temperature load during the 1590s and the 1690s, notable years of famines across Europe, as well as during the Great Frost of 1740 but to a lesser extent. All three periods were strongly linked to meteorological events\cite{baylor2015,Campbell_2016}, coinciding with extreme cold weather during the so-called 'Little Ice Age' \cite{littleicage,wanner2022variable}, accurately captured by the ModE-RA reconstruction in figures \ref{fig:key_temperatures_decades}. 

The famine of 1590–1598 was a major crisis across Europe\cite{clark1985european} and marked the last famine recorded in southern England \cite{alfani2017faminebook1}. Italy was particularly devastated, suffering what was possibly the worst famine in its history \cite{alfani2013calamitiesbook}. Most of Europe was affected, except for a few areas such as the Netherlands/Low Countries.  The famine probably affected southern Europe more severely in its initial phase (1590–93) and the centre and the north in the subsequent phase (1594–98)\cite{alfani2017faminebook}. This famine coincided with the lowest point of the Little Ice Age \cite{mann2009global,pages2013continental}.

The famine of 1693–1697 was especially severe in France, where approximately 1.5 million people died of starvation or disease between 1693 and 1694 \cite{dupaquier1690sfamine,alfani2017faminebook2}. In Italy, it was the second worst famine after the 1590s \cite{alfani2017faminebook3}. However, the regions most severely affected were in northern Europe, with the famine of 1696–1697 causing the death of 25–33\% of Finland's population and about 20\% of Estonia and Livonia's population \cite{alfani2017faminebook4}. There is documented evidence that the 1690s had some of the coldest winters recorded \cite{climatehisto_famines}. The Great Frost of 1740 \cite{greatfrost} led to a large Famine in Ireland \cite{engler2013irish,Dickson1997ArcticIT} but did not seem to trigger famines across the continent.

\begin{figure}
    \centering
    \includegraphics[width=0.6\linewidth]{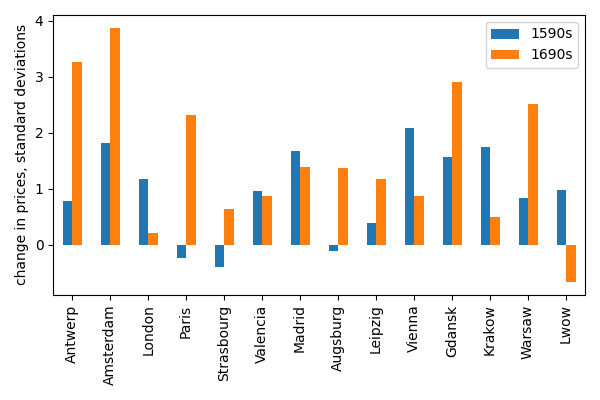}
    \caption{Change in standardized CPI in the 1590s (1589-1599) and the 1690s (1689-1699) }
    \label{fig:cpi_change_1590_1690}
\end{figure}

In preindustrial Europe, \citet{alfanitimingfamine} claim that high population pressure on resources was a common remote cause of a famine occurring, with crop-damaging weather being the proximate cause in most instances (usually persistent rain in the spring or more rarely, drought). \citet{climatehisto_famines} highlight the role of adverse weather as one of many factors in the emergence of famines. Our statistical analysis shows that cold temperatures in the mid to late 1590s (figure \ref{fig:sub1}) exacerbated the food crises (potentially even caused, in line with \citet{pfister2021climate}) which started in the South of Europe in the early 1590s (when temperatures were not particularly cold in the data). We make a similar claim for the 1690s famines that were exacerbated and even potentially caused by the extreme cold weather in the mid to late 1690s (see figure \ref{fig:sub2}).
We see an expected increase in prices across a vast majority of cities in both periods in figure \ref{fig:cpi_change_1590_1690}.

% \begin{figure}
%     \centering
%     \includesvg[width=0.99\linewidth]{weights_map_temp.svg}
%     \caption{Weights of the first canonical cariables for both Temperature and CPI matrices}
%     \label{fig:weights_map_temp}
% \end{figure}

\subsubsection{Prices, Droughts and Temperature}
We use SMCCA on the entire spatial fields (within latitudes -5 to 25 degrees and longitudes 40 to 60 degrees to stay within reasonable perimeter of our locations). This means our climate matrix becomes very high-dimensional as it has as many columns as there are latitude-longitude pairs,  hence the need for sparse methods designed to deal with such high-dimensional problems.

\begin{figure}
    \centering
    \includegraphics[width=0.8\linewidth]{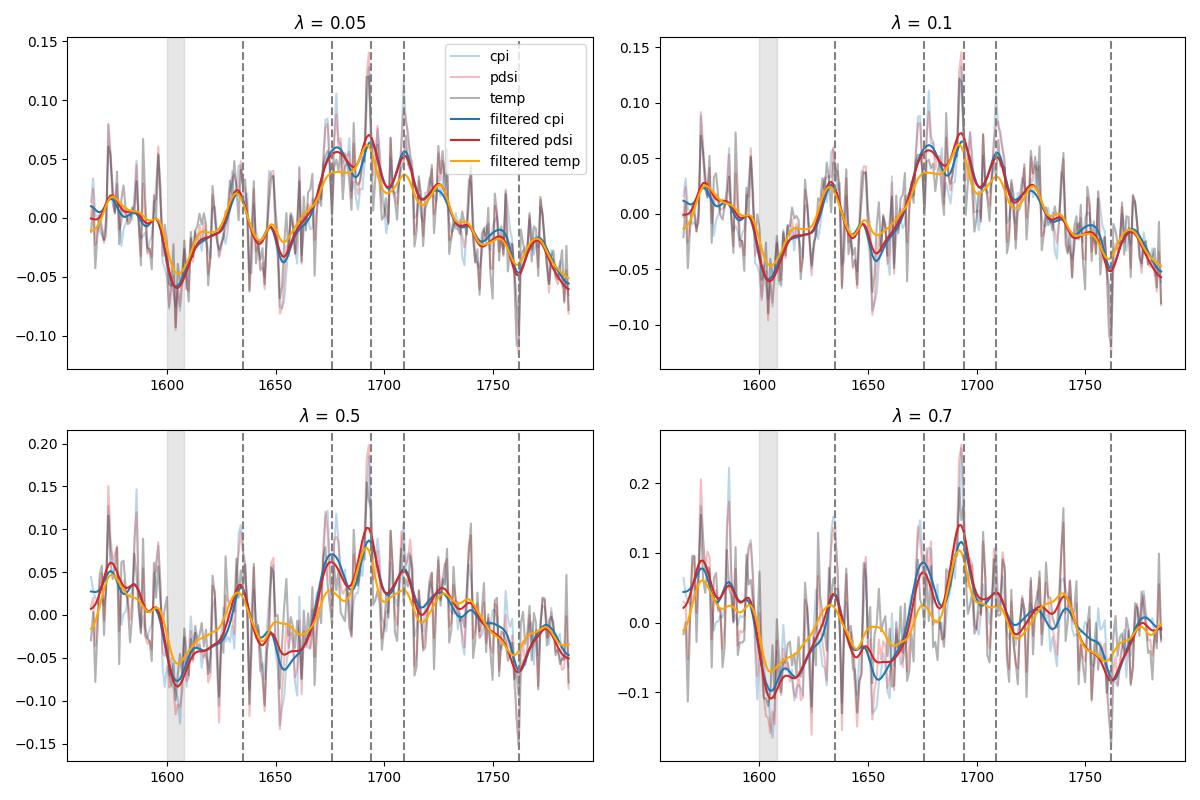}
    \caption{impact of regularization parameter $\lambda$ on canonical variables. Gray span covers 1600-1605, gray dashes are 1635, 1672, 1694, 1709, 1762}
    \label{fig:lambda_mcca}
\end{figure}

\begin{figure}
    \centering
    % Subfigure 1
    \begin{subfigure}[b]{0.99\linewidth}
       \caption{Temperature}
        \includegraphics[width=\linewidth]{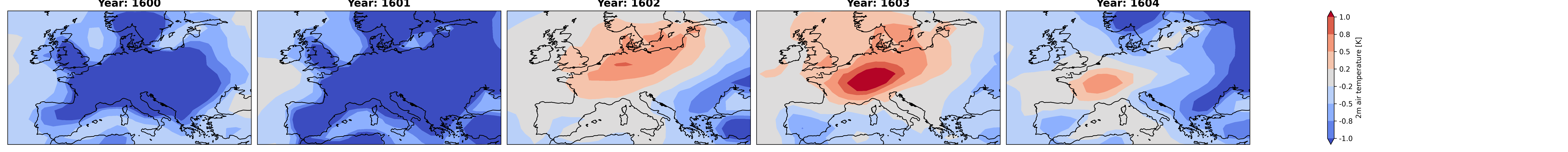}
     
        \label{fig:key_temperaturesMCCA_1600s}
    \end{subfigure}
    \hfill
    % Subfigure 2
    \begin{subfigure}[b]{0.99\linewidth}
            \caption{PDSI}
        \includegraphics[width=\linewidth]{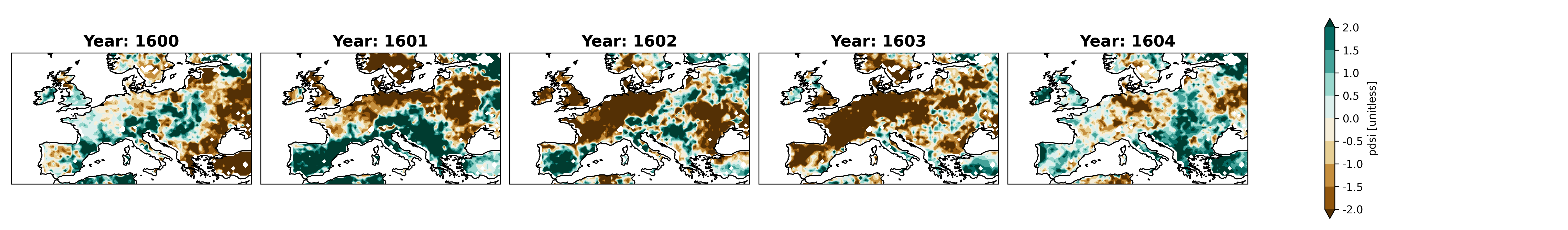}

        \label{fig:key_droughtsMCCA_1600s}
    \end{subfigure}

    \caption{Temperature and PDSI fields for early 1600s}
    \label{fig:mcca_fields_1600s}
\end{figure}

We first assess the influence of a common regularization parameter $\lambda$ on the first canonical variables in figure \ref{fig:lambda_mcca}, and conclude the results are robust to the choice of parameter. We now turn to the interpretation of the first canonical variables. A common set of years/periods are clearly singled-out every time: The early 1600s, 1635, 1672, 1694, 1709, 1762. We first turn to the early 1600s, which were marked by some severe episodes of drought and either very cold or very warm temperatures (see figure \ref{fig:mcca_fields_1600s}). This combination of changing temperatures and consecutive droughts likely triggered food shortages.

\begin{figure}
    \centering
    % Subfigure 1
    \begin{subfigure}[b]{0.99\linewidth}
       \caption{Temperature}
        \includegraphics[width=\linewidth]{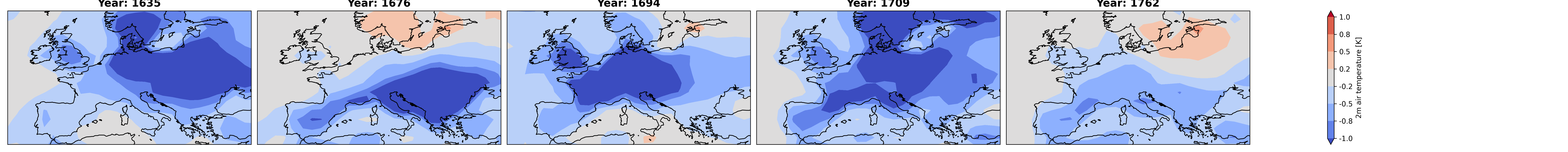}
     
        \label{fig:key_temperaturesMCCA}
    \end{subfigure}
    \hfill
    % Subfigure 2
    \begin{subfigure}[b]{0.99\linewidth}
            \caption{PDSI}
        \includegraphics[width=\linewidth]{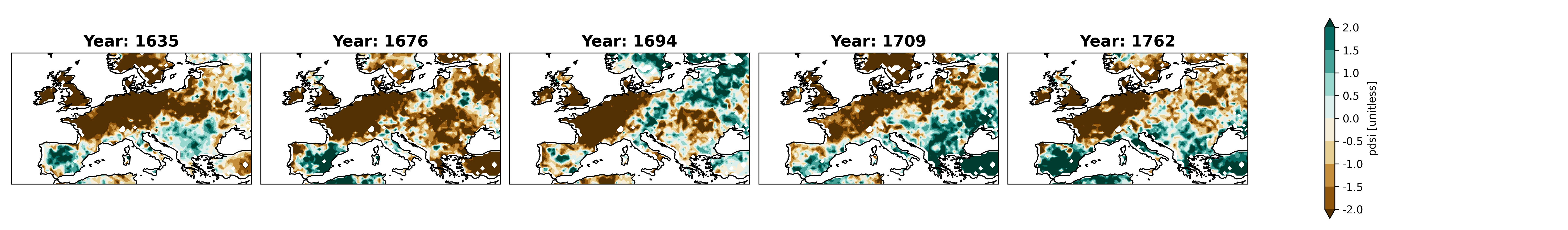}

        \label{fig:key_droughtsMCCA}
    \end{subfigure}

    \caption{Temperature and PDSI fields for years singled-out by the SmCCA}
    \label{fig:mcca_fields}
\end{figure}

We now look at the remaining years in figure \ref{fig:mcca_fields}. They were overwhelmingly characterized by severe droughts, some of which were notable like the 1694 drought \cite{owda}, the 1635 drought we already discussed, and by very cold temperatures, like during the Great Frost of 1709. Most of these years also fall within large-scale conflicts (the Thirty Years War for 1635, the War of Spanish Succession and the Great Northern War for 1709, and the Seven Years War for 1762), or famines (pan-European for the 1694 episode, 1601-1603\cite{alfani2017faminebook} for Nordic Europe, Italy, Ireland, and Low countries, France and Nordic countries for 1676\cite{Ljungqvist2023}). This reinforces the narrative that climate extremes aggravated the impact of societal  (wars, famines) on food prices. 

% As we are using the full grids for the SMCCA, we can look at the spatial loadings (ie canonical vectors) for each field (Temperature and PDSI) in figure \ref{fig:mcca_spatial_loadings}. We notice the PDSI loads on the Czech Lands while temperature loads on southern France (whose agriculture was very vulnerable to adverse climatic conditions, especially in 1693-94 where an estimated 1.5 million people died of starvation) and England. 

% \begin{figure}
%     \centering
%     % Subfigure 1
%     \begin{subfigure}[b]{0.49\linewidth}
%        \caption{PDSI}
%         \includesvg[width=\linewidth]{MCCA_weights_map_pdsi.svg}
     
%         \label{fig:MCCA_weights_map_pdsi}
%     \end{subfigure}
%     % Subfigure 2
%     \begin{subfigure}[b]{0.49\linewidth}
%             \caption{Temperature}
%         \includesvg[width=\linewidth]{MCCA_weights_map_temp.svg}

%         \label{fig:MCCA_weights_map_temp}
%     \end{subfigure}

%     \caption{Spatial loadings of first Canonical Vectors for the SmCCA, $\lambda=0.5$}
%     \label{fig:mcca_spatial_loadings}
% \end{figure}

% \input{regs_cor_temp}

\section{Discussion}

CCA methods are designed to capture latent patterns of correlation between sets of variables, but the interpretation of the latent factors has to be cross-referenced by domain knowledge. In our case, our findings are corroborated by both quantitative and qualitative studies, but we cannot make causality claims based on those, and have to remain conservative by talking about compounded effects of climate and societal change on food prices. 

Although we remain very cautious about some climate reconstructions used by previous work and chose to focus on temperature and droughts, further direction for this work will include adding reasonably robust reconstruction of other climate variables, potentially solar radiations, volcanic eruptions, soil moisture. A wider net of European cities would also help the mismatch in dimensions between the CPI fields and the higher resolution climate fields, but one would have to use grain prices with overlapping years, which will restrict the temporal resolution.

\section{Conclusion}

This study advances the literature on climate and economic history by applying robust multivariate statistical methods to explore the relationships between food prices, temperature, and drought conditions in early modern Europe. Utilizing econometric models and Canonical Correlation Analysis (CCA), along with its sparse variant (Sparse Multiple CCA), we identify significant correlations between climatic variables and food prices. Our findings highlight both the short-term and long-term impacts of climatic fluctuations, with a particular focus on the role of extreme weather events and their interactions with socio-political factors such as wars and famines.

The econometric analysis demonstrates a consistent negative relationship between temperature and food prices, corroborating prior findings that colder periods, often associated with the Little Ice Age, resulted in elevated prices. However, our models reveal limited evidence of a long-term, consistent relationship between drought conditions (as measured by the Palmer Drought Severity Index) and food prices. This suggests that while droughts were significant, their impact on prices may have been more context-dependent, varying by location and over time.

The CCA and Sparse Multiple CCA provide deeper insights into these relationships, allowing us to uncover complex, latent patterns of correlation. Our results show that significant climatic events, such as the Thirty Years' War, the cold spells of the 1590s and 1690s, and various severe droughts, were key drivers of economic disruptions. These findings strongly suggest that climatic shocks amplified economic stress, particularly in conjunction with other crises such as wars or famines, reinforcing the need to consider multiple, interacting stressors in historical economic analyses.

In conclusion, this study not only corroborates existing theories on the impact of climate on early modern European economies but also introduces a nuanced understanding of these dynamics by leveraging advanced statistical techniques. The findings contribute to a more sophisticated conceptualization of climate-economy interactions, emphasizing the importance of considering multivariate and multiscale relationships. Future research could expand upon these methodologies to include other environmental and social variables, further refining our understanding of historical climate-economic dynamics.

\bibliographystyle{unsrtnat}
\bibliography{references.bib}

\end{document}